# Full path single-shot imaging of femtosecond pulse collapse in air turbulence

I. Larkin[1], J. Griff-McMahon[1], A. Schweinsberg[2], A. Goffin[1], A.Valenzuela[2], and H. M. Milchberg[1*]

[1] Institute for Research in Electronics and Applied Physics, University of Maryland, College Park, MD 20742, USA

[2] CCDC Army Research Laboratory Aberdeen Proving Ground, MD 21005-5066, USA

[*]Corresponding author: milch@umd.edu

**In a single shot, we measure the full propagation path, including the evolution to pulse collapse, of a high power femtosecond laser pulse propagating in air. This technique enables single-shot examination of the effect of parameters that fluctuate on a shot-to-shot basis, such as pulse energy, pulse duration, and air turbulence-induced refractive index perturbations. We find that even in lab air over relatively short propagation distances, turbulence plays a significant role in determining the location of pulse collapse.**

The propagation of high peak power laser pulses through gases has applications spanning sub-millimeter scales for laser-driven relativistic electron acceleration [1] in thin gas jets to hundreds of meters in the atmosphere for applications in light detection and ranging (LIDAR) [2] and laser-induced breakdown spectroscopy (LIBS) [3]. In many cases, it is important to have a visualization of the full propagation path of the pulse in the gas. For long propagation ranges in the atmosphere, shot-to-shot variations from jitter in laser parameters and atmospheric fluctuations will lead to significant variations in the beam's transverse profile, axial energy deposition, and collapse location [4,5].

In prior work, records of long (> few cm) propagation profiles have been experimentally determined in several ways. One method is intercepting the beam along the propagation path and then, via propagation simulations, inferring aspects of the pulse propagation history to the point of interception [6,7]. Each shot, however, is sensitive to fluctuations and has a different propagation evolution. For femtosecond filaments, one approach for single-shot imaging is to use the recombination radiation from plasma generation [6]. However, the huge field of view needed to capture the full filament path precludes resolving axial detail. Another method is shot-by-shot scanning of a miniature microphone along the propagation path [8] to pick up the single-cycle cylindrical acoustic wave launched locally [9]. The acoustic signal is an excellent proxy for the local energy absorbed by the air, allowing a reconstruction of the laser pulse's axial energy deposition profile [8]. However, owing to unavoidable laser and air fluctuations, the full axial profile smooths over fluctuation-dependent details of interest by averaging many microphone traces at each position.

In this Letter, we apply a microphone array method to record, in a single-shot, the full axial energy deposition profile in air of a high peak power femtosecond pulse. We examine pulses that undergo optical collapse and then propagate as filaments. Our method enables visualization of the shot-to-shot dependence of filamentary propagation on fluctuations in laser parameters and on turbulence-induced air fluctuations.

In air, pulse collapse occurs due to positive self-lensing from nonlinear electronic and rotational contributions to the effective refractive index from nitrogen and oxygen. Collapse is arrested when the local laser intensity reaches the threshold for ionization of oxygen, $I_{th} \sim 5 \times 10^{13}$ W/cm$^2$ [10], after which the interplay between self-focusing and plasma defocusing leads to a self-guided beam whose central portion propagates as a tight, ~100 $\mu$m diameter "core" at intensity ~$I_{th}$ surrounded by a lower intensity periphery [11]. Most well-known applications of filaments [2,3,12-14] rely on well-controlled and reproducible propagation.

A femtosecond air filament deposits energy into its generated plasma channel and into the excitation of molecular rotational wavepackets in N2 and O2 [16, 17]. At each location along the filament path, the weakly ionized plasma recombines in less than ~10 ns [18], and the rotational excitation thermalizes on a ~100 ps timescale [19], leading to a very fast local increase in the thermal energy (and pressure) of neutral air [20]. The rise time of this pressure spike is much faster than the acoustic response timescale of the filament-heated gas ($\tau_a \sim a / c_s \sim 150$ ns, where $a \sim 50 \, \mu m$ is the filament core radius and $c_s \sim 3 \times 10^4$ cm/s is the sound speed in ambient air); the pressure spike drives an outwardly propagating single-cycle cylindrical acoustic wave [9,13] whose local axial amplitude is proportional to the local energy deposited. The peak in the acoustic signal registered by a microphone a distance $r$ from the filament at the axial position $z$ is $\Delta S_{mic}(z) \propto \Delta P(z)/\sqrt{r} \propto \Delta \epsilon(z)$, where $\Delta P(z)$ is the peak pressure amplitude of the acoustic wave and $\Delta \epsilon(z)$ is the laser energy per unit length absorbed at position $z$ through plasma generation and excitation of molecular rotation [8]. For this to be an accurate local measurement the microphone aperture width should be $w \lesssim 2r$, and $r \ll L$, where $L$ is the filament length.

The experimental setup is shown in Fig. 1. Filaments were generated by 1-5 mJ pulses at central wavelength $\lambda = 800$ nm from a Ti:Sapphire laser system, with pulse width adjusted in the range 40-200 fs by changing the pulse compressor grating spacing. The pulse energy was finely controlled in advance of the compressor with a motor-controlled $\lambda/2$ plate and a thin film polarizer (TFP). The pulse was then passed through a vacuum spatial filter to generate a near-Gaussian mode, after which a small portion was directed to a CCD camera calibrated with a power meter to measure the energy on every shot. After pulse compression, the transmission through a 95/5 (R/T) beamsplitter was sent to a single-shot autocorrelator to record the FWHM pulse duration on every shot, with the calibration adjusted for dispersion in the beamsplitter. The beam was then directed to a 4× down-collimating reflective telescope, with leakage from a dielectric turning mirror relay-imaged onto a Shack-Hartmann wavefront sensor (Imagine Optics HASO4 First) to determine the phase front curvature on every shot.

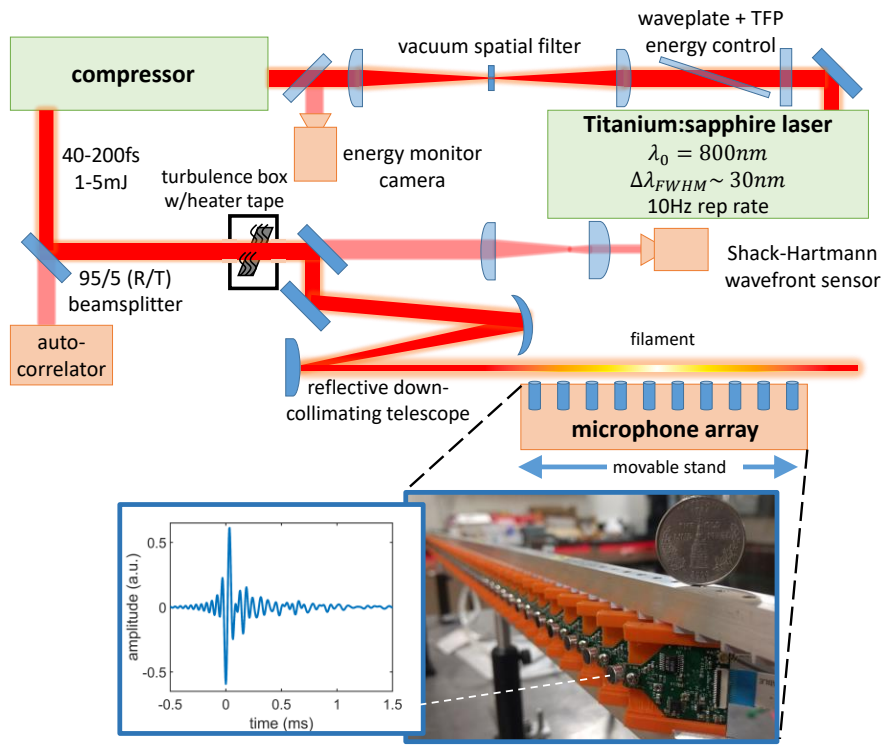

**Fig. 1.** Top: Optical setup. Bottom: Microphone array and sample single microphone signal, with white dashed line pointing to single microphone. A quarter is shown in the microphone array photo for scale.

The collimated beam emerged from the telescope with $w_0 = 1.65$ mm ($1/e^2$ intensity radius) and propagated along a 5.5 m run until a beam dump at the end of the lab. The collapse was governed entirely by nonlinear phase accumulation in air without any assistance of linear beam focusing; even a weak lens can stabilize the collapse position, as shown later.

The microphone array was mounted on a mobile cart and aligned to be at a fixed radial position of ~3 mm from the filament. The array is composed of 64 miniature electret condenser microphones (Panasonic WM-61A) with aperture $w = 6$ mm, each mounted to a separate circuit board that does 24 bit A/D conversion at 44.1 kHz. The microphone, with a peak frequency response of ~20 kHz, registers the single-cycle acoustic wave (where $\tau_a^{-1} \sim$ 3 MHz) as an impulse response. The microphones were mounted at 2 cm longitudinal spacing for a full span of 126 cm, providing sufficient axial resolution to capture details of the shot-to-shot variations along the full filament extent. The microphones are connected to a central hub with FPGA synchronization of the array and USB data transfer to a computer. Data was acquired continuously from the microphones and at 10 Hz from all other diagnostics.

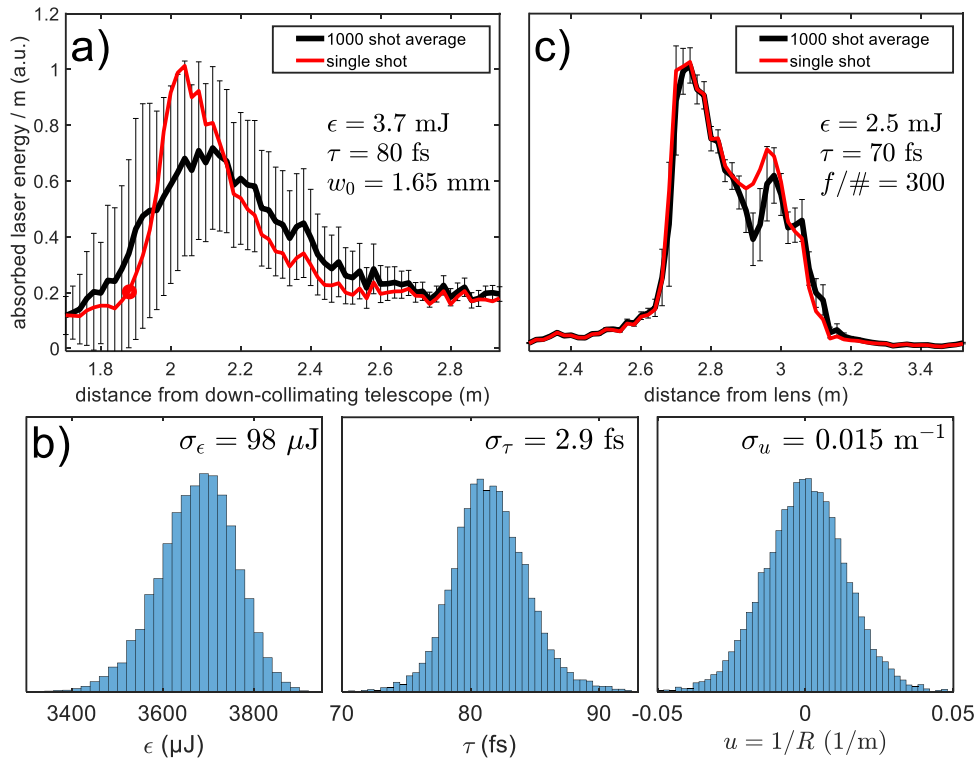

**Fig. 2.** Microphone array signal from a single-shot (red curve) compared to a 1000 shot average (black curve). The pulse propagates left to right. Error bars associated with the average curves are the $\pm$ standard deviations of the signals at each microphone. (a) Filamentation with collimated beam. Laser: 3.7 mJ, 80 fs FWHM, $w_0 = 1.65$ mm. (b) Histograms of pulse energy ($\varepsilon$), FWHM pulse duration ($\tau$) and wavefront curvature ($u = 1/R$) over 10,000 shots, with standard deviations shown in the panels. (c) Lens-assisted filamentation at $f/300$. Laser: 2.5 mJ, 70 fs FWHM, $w_0 = 5$ mm incident on 3 m lens.

For a collimated beam that collapses and propagates as a filament, there can be large shot-to-shot fluctuations in the axial energy deposition profile. Figure 2(a) compares a single-shot array trace, where each point corresponds to the peak of the microphone signal at that location (a sample microphone trace is shown in Fig. 1), to a trace averaged over 1000 shots at 10 Hz. The pulse propagates from left to right in the plots. The interval between shots is much greater than the ~ 2 ms needed for the air density to recover [9, 13]. The error bars associated with the average are the ± standard deviations of the signals at each of the array microphones. The single-shot (red) trace shows a much sharper increase to its maximum than the average trace. This is because the single-shot trace has captured the onset of pulse collapse and filamentation, whereas the average trace has smeared this region out. By contrast, Fig. 2(c) shows that even relatively weak assistance by a lens, here at $f/300$, greatly stabilizes the filament onset location on the left, where the single-shot and average curves largely track one another except for some deviation near the second hump of the curves. The error bars on the average curve are much smaller, in agreement with the single microphone axial scans of $f/600$ lens-assisted filaments in [8].

While the microphone array provides the energy deposition profile over the full filamentary propagation path, we now concentrate on the pulse collapse and filament onset location, as this is a measure of the path-integrated effect of the fluctuations leading to the shot-to-shot nonlinear propagation variations. These fluctuations are either intrinsic to the pulse (energy, pulse width, phase front) or imposed on the pulse (phase front perturbations by externally supplied turbulence). To proceed, we define the collapse location as the axial position where energy deposition per unit length reaches 20% of its peak value. This location is marked as filled-in circle on the single-shot (red) trace in Fig. 2(a); 20% and higher gives very similar results, while thresholds as low as 10% result in some non-collapse locations being counted.

The intrinsic pulse fluctuations are measured, as shown in Fig. 1, with the pulse energy monitor (pulse energy $\varepsilon$), single-shot autocorrelator (pulse FWHM $\tau$), and wavefront sensor (wavefront radius of curvature $R$). For the wavefront measurement we consider, as intrinsic to the laser, fluctuations measured at the down-collimating telescope. As discussed below, the wavefront fluctuations are from air turbulence over the ~10 m propagation path from the spatial filter to the telescope, plus a much smaller contribution making it through the spatial filter from upstream in the laser system. The externally-imposed fluctuations are from controlled turbulence introduced into the beam by a length of heater tape enclosed at the base of a 15 cm × 15 cm box with an open top and small apertures for the laser beam to enter and exit ~6 cm above the tape. The turbulence box was positioned immediately preceding the reflective telescope, as shown in Fig. 1.

The turbulence strength in the lab air or imposed by the box was measured using a spatially filtered λ = 532 nm CW diode probe laser sampled at 10 Hz by the 1 ms electronic shutter of a CCD camera. Given uniform turbulence along a propagation distance $L$, the spatial deflection of a laser beam has the variance $\sigma^2 = 0.97\, C_n^2 D^{-1/3} L^3$ [21], where $\sigma$ is the standard deviation of the beam centroid on a camera, D is the average FWHM beam diameter (all in meters), and $C_n^2$ is the refractive index structure parameter. For propagation of the probe laser across the lab under typical conditions, with all laser power supplies (a heat source) running, we obtained $C_n^2 = 6.4 \times 10^{-14}$ m$^{-2/3}$.

For the fixed nominal laser parameters of Fig. 2(a) ($\varepsilon = 3.7$ mJ, $\tau = 80$ fs, and $w_0 = 1.65$ mm at the output of the down-collimating telescope) without the turbulence box, Fig. 2(b) shows histograms of the fluctuations over 10,000 shots, with the standard deviations $\sigma_\varepsilon$, $\sigma_\tau$ and $\sigma_u$ shown in the panels, where $u = 1/R$. As the distributions are symmetric about their peaks, and these parameters are uncorrelated with one another, we take the fluctuations to be random. These small relative fluctuations can be considered as independently affecting the pulse collapse location $z_{cl}$, whose standard deviation can then be written as $\sigma_{z_{cl}} = [\sigma_\varepsilon^2 (\partial z_{cl}/\partial \varepsilon)^2 + \sigma_\tau^2 (\partial z_{cl}/\partial \tau)^2 + \sigma_u^2 (\partial z_{cl}/\partial u)^2]^{1/2}$, with the gradient $(\partial/\partial \varepsilon, \partial/\partial \tau, \partial/\partial u) z_{cl}$ determined from many shots measuring $z_{cl}$ vs. $(\varepsilon, \tau, u)$ and evaluated at the mean point $(\bar{\varepsilon}, \bar{\tau}, \bar{u})$. The expression for $\sigma_{z_{cl}}$ can be used to isolate the contributions of each of the fluctuating variables on the pulse collapse location. To do this, the nominal pulse energy or pulse duration are fixed while the other parameter is scanned beyond the standard deviations ($\sigma_\varepsilon$ or $\sigma_\tau$) to determine the value of the partial derivatives more accurately than from the points clustered within the standard deviations. For the wavefront curvature, independent control of $u$ was experimentally difficult; however $\sigma_u$ was sufficiently wide to provide a reasonable value for $\partial z_{cl}/\partial u$.

The results of these scans are shown in Fig. 3, where the green lines are a second order polynomial least squares fit whose local slope, through the mean values $\bar{\varepsilon}, \bar{\tau},$ and $\bar{u}$, is shown in the panels. The vertical spread in $z_{cl}$ points, for example in Fig. 3(a) for a fixed value of $\varepsilon$, is reflective of the random fluctuations of $\tau$ and $u$. Similarly, the spreads in $z_{cl}$ in Fig. 3(b) and in Fig. 3(c) are reflective of fluctuations in $(\varepsilon, u)$ and $(\varepsilon, \tau)$, respectively. Importantly, the vertical spread appears roughly constant within each of the panels of Fig. 3. This further reflects the lack of correlation among the parameters.

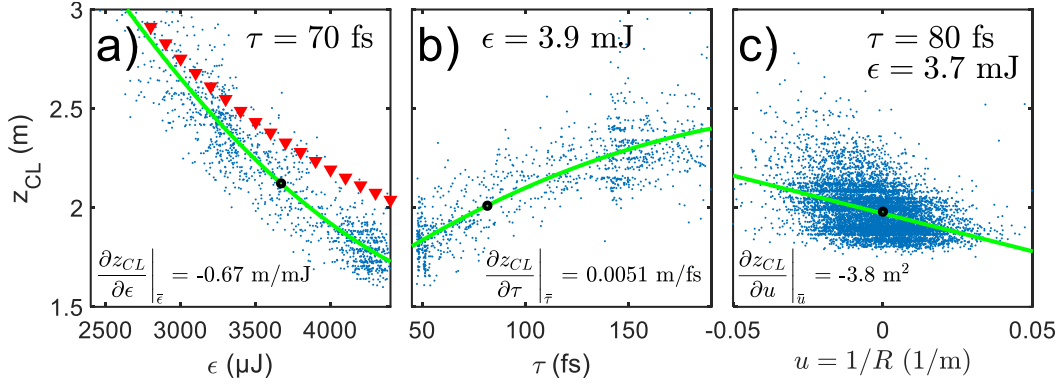

Fig. 3. (a) Collapse location (from down-collimating telescope) vs. scanned pulse energy and (b) pulse duration. (c) Collapse location vs. intrinsic fluctuations in wavefront curvature. Green curves: quadratic best fits to data points. The black circles indicate the mean values $\bar{\varepsilon}, \bar{\tau},$ and $\bar{u}$. The red triangles in (a) are from an UPPE simulation without turbulence.

For the collimated beam of Fig. 2(a) and 2(b), using $\sigma_\varepsilon = 98$ µJ and $(\partial z_{cl}/\partial \varepsilon)_{\bar\varepsilon} = -0.67$ m/mJ from Fig. 3(a), we expect that pulse energy fluctuations alone would give a standard deviation of collapse location $\sigma_{z_{cl}}^\varepsilon \sim \sigma_\varepsilon |(\partial z_{cl}/\partial \varepsilon)_{\bar\varepsilon}| \sim 6.6$ cm. Similarly, for pulse width variations alone, $\sigma_{z_{cl}}^\tau \sim \sigma_\tau |(\partial z_{cl}/\partial \tau)_{\bar\tau}| \sim 1.5$ cm and for intrinsic phase front fluctuations, $\sigma_{z_{cl}}^u \sim \sigma_u |(\partial z_{cl}/\partial u)_{\bar u}| \sim 5.7$ cm. Therefore, energy and phase front fluctuations are much more important than fluctuations in the pulse duration for determining the repeatability of filaments. It is evident that air turbulence, even indoors in a lab, is an important factor over distances as short as a few meters.

We modeled the experiment of Fig. 2 using a 2D+1 (2D space plus time) GPU-based simulation using the unidirectional pulse propagation equation (UPPE) [22, 23], which includes the Kerr and molecular rotational response of air, plus ionization. To model the effect of energy variations alone, these simulations do not include turbulence. The resulting simulation points (red triangles) are overlaid in Fig. 3(a), showing good agreement with the measurements, especially at lower pulse energy. To simulate turbulence (see later), we used a modified von Karman spectrum seeded with random noise and inverse Fourier transformed to generate 2D phase screens [24] every centimeter.

To determine the contribution to wavefront fluctuations of beam propagation before the spatial filter, where the beam was enclosed by a sealed box inside a HEPA tent, the wavefront sensor was placed immediately following the filter. Based on a 10,000 shot sample, we measured $\sigma_{z_{cl}}^u = 0.0042$ m$^{-1}$, which we attribute mainly to fluctuations in thermal lensing in the laser rods. This spread is more than 3× smaller than $\sigma_{z_{cl}}^u \sim 0.015$ m$^{-1}$ at the down-collimator.

Further exploring the effects of turbulence on femtosecond pulse collapse, we employed the turbulence box preceding the reflective telescope, as shown in Fig. 1. For this experiment, the laser parameters were $\varepsilon$ =2.8–2.9 mJ and $\tau$ =45 fs. Here, since stronger turbulence was localized to a short axial region, its strength was determined using $C_n^2 L = \theta^2 D^{1/3}/2.91$ [25], which depends on the angular deflection $\theta$ (rad) of the $\lambda$=532 nm probe beam, the average beam diameter $D$, and the length $L$ (= 15 cm) of the turbulence box. Figure 4(a) plots $C_n^2$ versus heater voltage, while Fig. 4(b) plots $\sigma_{z_{cl}}$ and mean collapse location $\overline{z_{cl}}$ versus $C_n^2 L$. For each heater tape voltage in this experiment, several (3-5) 1000 frame (each 1 ms) sets were taken to determine $C_n^2 L$, and seven sets of 1000 filament shots were taken. These multiple sets were taken to cover potential laser and environmental drifts during our runs. As seen in Fig. 4(a), over the voltage scan 10-100 V, $C_n^2 L$ ranged from $2.7 \times 10^{-13}$ m$^{1/3}$ to $1 \times 10^{-11}$ m$^{1/3}$. Figure 4(b) plots the standard deviation of the collapse location and the mean collapse location, where it is seen that $\sigma_{z_{cl}}$ increases with $C_n^2 L$ but $\overline{z_{cl}}$ is roughly constant. Histograms of $z_{cl}$ are plotted in Fig. 4(c), showing the increased spread in collapse location for increased heater voltage and turbulence. We modeled the collapse variability for 60 V with a run of 50 3D+1 UPPE simulations including room turbulence ($C_n^2 L \sim 6.4 \times 10^{-13}$ m$^{1/3}$ for the $L \sim 10$ m propagation path from the spatial filter to the telescope) and the turbulence box ($C_n^2 L \sim 2.4 \times 10^{-12}$ m$^{1/3}$ over 15 cm), giving $\sigma_{z_{cl}}^{sim} = 19$ cm, in good agreement with $\sigma_{z_{cl}} = 21.5$ cm at 60 V in Fig. 4(c).

We note that the propagation simulations in [4] predict a reduction in $\overline{z_{cl}}$ with increased turbulence, in contrast to our experiments and simulations. However, the beams simulated in [4] are wider than the turbulence inner scale of ~1 mm, and are more susceptible to the modulation instability than our $w_0 \sim 1.7$ mm beams.

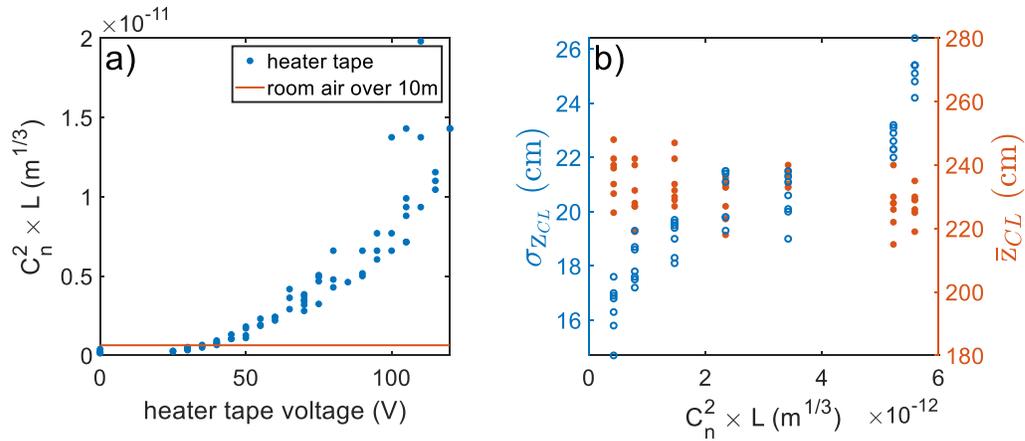

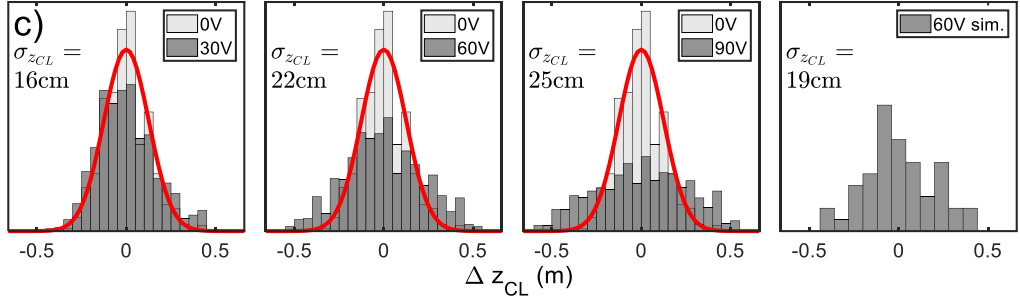

Fig. 4. (a) $C_n^2 L$ measured in turbulence box (using a λ = 532 nm probe) vs. heater tape voltage. The brown baseline is the room turbulence level. (b) Mean collapse location $\overline{z_{cl}}$ and standard deviation $\sigma_{z_{cl}}$ vs. turbulence strength. (c) Histograms of collapse location $\Delta z_{cl}$ relative to $\overline{z_{cl}}$ vs. heater tape voltage. The histogram for each voltage overlays the V = 0 histogram, whose best-fit Gaussian is the red curve. Right panel: Histogram for 50 3D+1 UPPE simulations for 60 V.

To summarize, we have performed measurements recording the full axial energy deposition profile of a nonlinearly propagating laser pulse over macroscopic laboratory distances in a single shot, using a linear array of synchronized microphones. In particular, we have examined the sensitivity of pulse collapse of high peak power femtosecond pulses to fluctuations in pulse width, pulse energy, and wavefront curvature. We have found that pulse energy and room air turbulence-induced wavefront curvature fluctuations are the dominant contributions. The important role of air turbulence, even over relatively short distances in the laboratory, is confirmed through 3D + 1 propagation simulations.

The authors thank Eric Rosenthal and Sina Zahedpour for useful discussions. This work was supported by AFOSR (FA9550-16-1-0284 & FA9550-16-1-0121), ARO (W911NF-14-1-0372 and W911NF-16-2-0233), and ONR (N00014-17-1-2705 & N00014-17-1-2778)